 \documentclass[%
twocolumn,
superscriptaddress,
amsmath,amssymb,
aps,
prl
]{revtex4-1}

\usepackage{color}
\usepackage{graphicx}

\begin{document}

\title{Light stops at exceptional points}

\author{Tamar Goldzak}
\affiliation{Schulich Faculty of Chemistry and Faculty of Physics, Technion -- Israel Institute of Technology, Haifa, 32000, Israel}
\author{Alexei A. Mailybaev}\email{alexei@impa.br}\affiliation{Instituto Nacional de Matem\'atica Pura e Aplicada -- IMPA, 22460-320 Rio de Janeiro, Brazil}
\author{Nimrod Moiseyev}\email{nimrod@tx.technion.ac.il}\affiliation{Schulich Faculty of Chemistry and Faculty of Physics, Technion -- Israel Institute of Technology, Haifa, 32000, Israel}

\begin{abstract}
Almost twenty years ago the light was slowed down to less than $10^{-7}$ of its vacuum speed in a cloud of ultracold atoms of sodium. Upon a sudden turn-off of the coupling laser, a slow light pulse can be imprinted on  cold atoms such that it can be read out and converted into photon again. In this process, the light is stopped by absorbing it and storing its shape within the atomic ensemble. Alternatively, the light can be stopped at the band edge in photonic-crystal waveguides, where the group speed vanishes. Here we extend the phenomenon of stopped light to the new field of parity-time (PT) symmetric systems. 
We show that zero group speed in PT symmetric optical waveguides can be achieved if the system is prepared at an exceptional point, where two optical modes coalesce. This effect can be tuned for optical pulses in a wide range of frequencies and bandwidths, as we demonstrate in a system of coupled waveguides with gain and loss.
\end{abstract}

\maketitle

Electromagnetically induced transparency and other techniques for controlling atom-light interactions developed in last decades~\cite{bigelow2003superluminal,*fleischhauer2005electromagnetically,baba2008slow} affect drastically the form of light propagation~\cite{wang2000gain,ginsberg2007coherent,*everett2016dynamical}, e.g., by reducing a group speed to just several meters per second~\cite{hau1999light}. Turning off a control field, one dissipates the optical energy but stores a signature of the slow light pulse in atomic system~\cite{liu2001observation,heinze2013stopped}. Alternatively, with two counter-propagating control fields one can preserve some optical energy during the storage in the form reminiscent of a standing wave~\cite{bajcsy2003stationary,ruter2010observation}.
However, the resonant mechanism of this phenomenon imposes substantial limitations on the operation frequencies and signal bandwidth.
Another mechanism of stopped light was proposed in photonic crystals (materials with periodic dielectric constants~\cite{joannopoulos1997photonic}) at the photonic band edge~\cite{yanik2004stopping,*yanik2004stoppingB,baba2008slow}.

In this work, we disclose the relation of the stopped light effect with the phenomenon of exceptional point (EP). The latter characterizes a system, in which two propagating modes coalesce both in their frequencies and eigenvectors. The appearance of EP is only possible in non-Hermitian systems~\cite{moiseyev2011non}, and it triggers a number of exciting phenomena discovered recently across different areas of physics~\cite{berry2004physics,bender2007making,*uzdin2011observability,*heiss2012physics,*peng2014loss,*xu2016topological,*doppler2016dynamically}. We show that the group velocity vanishes, i.e., a light pulse is fully stopped if the waveguide (WG) is designed exactly at the EP. This result opens conceptually new possibilities for designing slow light devices, which exploit generic properties of the EP and, therefore, may offer much larger freedom for technical implementation and operational capability. 

With the described EP mechanism, we extend the stopped-light phenomenon to the new field of PT-symmetric systems. We show that the controllable full-stop of a light pulse is possible in a PT-symmetric WG, which is characterized by the loss accurately balanced with the gain. The fundamental symmetries of parity and time have recently being exploited to enable the spatial guiding and selection of propagating radiation, that could ultimately underpin a new generation of sophisticated, integrated photonic devices~\cite{makris2008beam,*ruter2010observation,*kottos2010optical}.
 PT-symmetric WGs are known to have constant-amplitude propagating modes (real spectra) when the gain/loss parameter does not exceed a problem dependent critical value~\cite{klaiman2008visualization}. 
We focus on a conceptual design for an experiment, where a signal can be fully stopped by changing the gain/loss parameter adiabatically to the value at EP.

Let us first consider a two-dimensional WG with the propagation axis $z$. Transverse electric modes are described by scalar complex states $\phi = \psi(x)\, e^{i\beta z-i\omega t}$ satisfying the Helmholtz equation (derived from the full Maxwell equations) as~\cite{jackson}
\begin{equation}
\frac{\partial^2\psi}{\partial x^2}
+\left(\frac{n^2\omega^2}{c^2}-\beta^2\right)\psi
= 0,
\label{eqR1}
\end{equation}
where $\omega$ is the frequency, $\beta$ is the propagation constant, $n(x,\omega)$ is the refractive index and $c$ is the light speed. In the media with gain and loss, the refractive index is complex: a positive imaginary part of $n$ indicates the region with loss, while negative imaginary part describes the gain. In a PT-symmetric WGs, the gain and loss are balanced and satisfy the condition~\cite{klaiman2008visualization}:
$n(x) = n^*(-x)$, where the complex conjugation (star) corresponds to the time reversal that interchanges the gain and loss. Equation (\ref{eqR1}) is equivalent to the one-dimensional stationary Schr\"odinger equation for the complex (non-Hermitian) potential, with the propagating modes corresponding to bound states~\cite{moiseyev2011non}.

As soon as the strength of gain and loss is below a problem dependent critical value~\cite{klaiman2008visualization}, non-decaying modes exist with real propagating constants $\beta$ and complex PT-symmetric eigenfunctions, $\psi(x) = \psi^*(-x)$.
Thus, the phase speed of each mode is defined as $v_p = \omega/\beta$, while the group speed is $v_g = \left(d\beta/d\omega\right)^{-1}$. 
For a nondegenerate bound-state solution, differentiating equation (\ref{eqR1}) with respect to $\omega$ yields
\begin{equation}
\left(\frac{\partial^2}{\partial x^2}
+\frac{n^2\omega^2}{c^2}-\beta^2\right)\frac{\partial\psi}{\partial \omega}
+\left(\frac{\partial(n^2\omega^2/c^2)}{\partial\omega}-\frac{\partial \beta^2}{\partial \omega}\right)\psi
= 0.
\label{eqM1}
\end{equation}
Following the classical perturbation theory~\cite{landauquantum,moiseyev2011non}, one multiplies this expression by $\psi(x)$ and integrates with respect to $x$. The terms with $\partial\psi/\partial\omega$ cancel in the resulting expression after integrating by parts and using (\ref{eqR1}). The remaining terms yield expression for the group speed as
\begin{equation}
v_g = \left(d\beta/d\omega\right)^{-1}
= \frac{2c^2\beta\int \psi^2dx}{\int [\partial (n^2\omega^2)/\partial \omega]\psi^2dx}
\label{eqR2}
\end{equation}
%For comparison, we recall the group velocity $v_g = c\left(n+\omega\frac{dn}{d\omega}\right)^{-1}$ in a homogeneous medium.

In the perturbation theory for non-Hermitian operators, one must distinguish between the right and left complex eigenfunctions, which in our case are given by $\psi$ and $\psi^*$, respectively.
Thus, the nominator in (\ref{eqR2}) represents the so-called c-product, $\langle \psi^*|\psi\rangle = \int \psi^2 dx$, \textit{with no complex conjugation}, while the PT-symmetry ensures that the full integral is real but not necessarily positive. Hence, as it follows from equation~(\ref{eqR2}), the group speed vanishes if and only if
$\int \psi^2dx = 0$,
provided that the integral in the denominator is nonzero. The latter condition is generic and can be easily verified in each specific problem. The c-product self-orthogonality of the propagating mode is the well-known condition for the exceptional point (EP), which is a branch point singularity \cite{moiseyev2011non}. At the EP, two propagating modes coalesce both in propagation constants $\beta$ and  corresponding functions $\psi(x)$. This proves the main result of our work demonstrating that \textit{the group speed in a PT-symmetric WG vanishes at (and only at) an exceptional point}.

At the EP, the local expansion of eigenvalues starts with a square root term~\cite{seyranian2003multiparameter,moiseyev2011non},
$\beta-\beta_{EP} \propto \sqrt{\omega-\omega_{EP}}$,
which implies that
$d\beta/d\omega = \infty$ and $v_g = \left(d\beta/d\omega\right)^{-1} = 0$ at the EP. 
This provides a simple explanation of our phenomenon, because any system having a divergence in the dispersion curve (large derivative $d\beta/d\omega$) will exhibit slowing in the group velocity. This argument relies exclusively on the presence of the EP, with no reference of the PT-symmetry. The problem is that in conventional systems this effect will also lead to losses. The balance between gain and loss in a PT symmetric system eliminates this problem: the light intensity remains constant before reaching the EP.
Also, the real spectrum of the PT symmetric system simplifies a definition of the group speed, which is a nontrivial issue for a general system with gain and loss. The direct link between the EP and zero group speed makes the proposed effect robust to various imperfections, as the proximity to the EP can be effectively controlled by tuning two arbitrarily chosen parameters of the system~\cite{seyranian2003multiparameter,moiseyev2011non}.

A specific device with desired properties can be constructed by attaching layers of materials with different indices of refraction. The refractive index can be engineered, e.g., via the photorefractive nonlinearity or effective index as in metamaterials, while the spectrum of gain/loss can be engineered by using quantum well structures.
 The PT-symmetry is achieved if one gain guided mode (negative $\mathrm{Im}\,n$) couples with an exactly balanced loss guided mode (positive $\mathrm{Im}\,n$)~\cite{siegman2003propagating,klaiman2008visualization}, with a profile of the refractive index shown in Fig.~1. Describing the effective light intensities of these modes by two complex variables $(\psi_1,\psi_2)$, one obtains a simple model in the form of $2\times 2$ PT-symmetric non-Hermitian system
\begin{equation}
\left(\begin{array}{cc}\beta_w^2-i\alpha k & \delta \\
\delta & \beta_w^2+i\alpha k \end{array}\right)
\left(\begin{array}{c}\psi_1 \\ \psi_2\end{array}\right) = \beta^2\left(\begin{array}{c}\psi_1 \\ \psi_2\end{array}\right).
\label{eqR3}
\end{equation}
Here $\beta_w = n_wk$ is the real propagation constant of each separate WG with the effective index of refraction $n_w$ and $k = \omega/c$, $\delta$ describes the coupling, and $\alpha$ determines the gain in one WG and the loss in the other. The system with no gain/loss ($\alpha = 0$) has one symmetric and one antisymmetric mode, with $\beta^2 = \beta_w^2\pm\delta$ and $(\psi_1,\psi_2) = (\pm1,\,1)$. When gain and loss are taken into account, one finds $\beta^2 =  \beta_w^2\pm\sqrt{\delta^2-\alpha^2k^2}$. With increasing $\alpha$, the real propagating constants come closer and coalesce at the EP given by $\alpha_{EP} = \delta/k$. The corresponding two eigenvectors  coalesce too, with the resulting vector $(\psi_1,\psi_2) = (1, i)$ satisfying the self-orthogonality condition $\psi_1^2+\psi_2^2 = 0$.

\begin{figure}
\centering
\includegraphics[width=0.9\columnwidth]{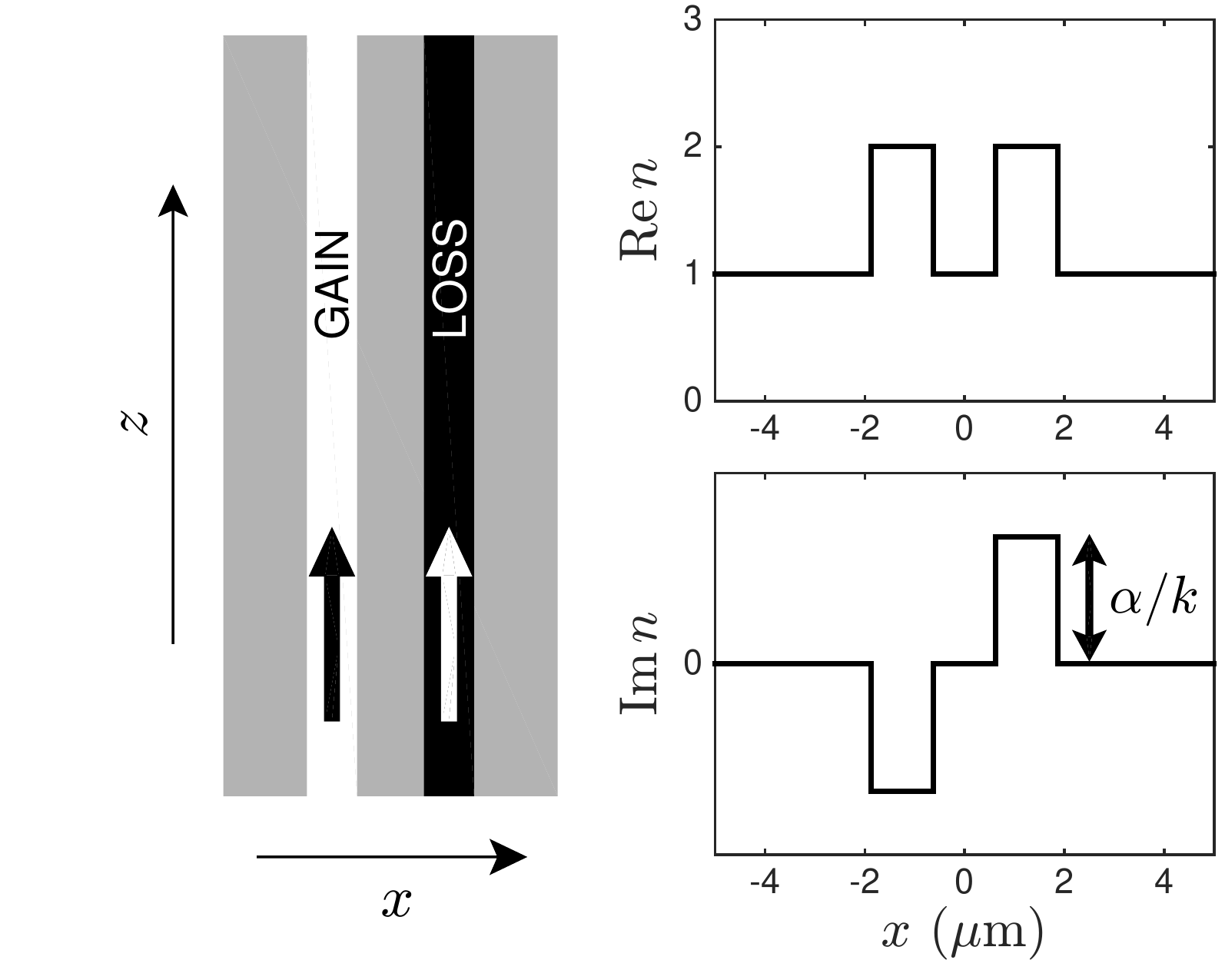}
\caption{Coupling between the gain guided mode and the loss guided mode provides a PT-symmetric system with the refractive index profile such that $n(x)=n^*(-x)$. A control parameter $\alpha$ defines the gain/loss strength in the WGs as $\mathrm{Im}\, n = \pm \alpha/k$ with $k = \omega/c$. In simulations of Figs.~3(a) and 4, we use the WGs of width $W = 1.25\,\mu m$ and the distance between them $D = 1.25\,\mu m$. Analogous setup \cite{klaiman2008visualization} with different parameters has been used in the first experiment \cite{ruter2010observation}, which demonstrated the PT-symmetry in optics.}
\label{Figure1}
\end{figure}

The full-stop of a Gaussian pulse can be accomplished by an adiabatic increase of the gain/loss parameter to the value $\alpha_{EP}$, as it was also proposed in the context of photonic-crystal waveguides~\cite{yanik2004stoppingB,tanaka2007dynamic}. 
Varying the gain/loss parameter in our non-Hermitian system would be best done via parametric nonlinear gain, which separates the variation in the gain/loss from affecting the real part of refractive index, avoiding restrictions imposed by the Kramers--Kronig relations. 
Nonlinear parametric interactions operating at ultrafast rates~\cite{weiner2011ultrafast} can be engineered using synchronously-pumped optical parametric oscillators, where the nonlinear medium is in a cavity and pumped with a pulse at repetition rate matched to cavity, or optical parametric amplifiers pumped without a cavity by femtosecond pulse. Usually these utilize $\chi^{(2)}$ crystals, which are commercial technologies. Another choice is $\chi^{(3)}$ materials, through non-degenerate four-wave-mixing interactions, where the pumps serve as gain/loss for the signal beams~\cite{agrawal2013nonlinear}.

Time-dependent solutions for the simplified model (\ref{eqR3}) can be found using the system of coupled wave equations
\begin{equation}
\begin{array}{l}
\displaystyle
\frac{n_w^2}{c^2}\frac{\partial^2\phi_1}{\partial t^2}
-\frac{\alpha}{c}\frac{\partial\phi_1}{\partial t}-\delta\,\phi_2
-\frac{\partial^2\phi_1}{\partial z^2} = 0,\\[12pt]
\displaystyle
\frac{n_w^2}{c^2}\frac{\partial^2\phi_2}{\partial t^2}
+\frac{\alpha}{c}\frac{\partial\phi_2}{\partial t}-\delta\,\phi_1
-\frac{\partial^2\phi_2}{\partial z^2} = 0.
\end{array}
\label{eqM2}
\end{equation}
It is straightforward to check that this system is equivalent to equation (\ref{eqR3}) for a single-mode solution
$(\phi_1,\phi_2) = (\psi_1,\psi_2)\,e^{i\beta z-i\omega t}$. Furthermore, it is easy to see that the model is PT-symmetric under the transformation: $\phi_1(z,t) \to \phi_2(-z,-t)$ and $\phi_2(z,t) \to \phi_1(-z,-t)$. System~(\ref{eqM2}) was simulated numerically using the pseudo-spectral method in a large periodic domain. Initial condition at $t = 0$ was taken in the form of a Gaussian pulse corresponding to the antisymmetric mode of the system with no gain and loss, $(\phi_1,\phi_2) = (-1,1)\,A\int
\exp\left(-\frac{(\beta-\beta_0)^2}{2\sigma^2}+i\beta z\right)d\beta$, for the mean propagation constant $\beta_0 = 0.8$, standard deviation $\sigma = 0.01$ and arbitrary prefactor $A$. This value $\beta_0 = 0.8$ corresponds to the EP at the final time when $\alpha_{EP} = 1$, see Fig.~2(a).
In simulations, we used a finite window $0.75 \le \beta \le 0.85$ to avoid instabilities, which occur for some propagation constants outside this interval. In practical applications, such instabilities (if they appear) must be suppressed for efficient operation of the system.

Numerical simulation of such time-dependent dynamics with the model (\ref{eqM2}) is presented in Fig.~2, where a Gaussian pulse is prepared initially in the antisymmetric mode of the system with no gain and loss.
In full agreement with our theoretical prediction, with the increase of the gain/loss parameter in time, the pulse slows down and stops at the EP (graphs at latest times collapsed to a single curve). A backward change of the gain/loss parameter brings the signal to its original mobile form.

\begin{figure}
\centering
\includegraphics[width = 1\columnwidth]{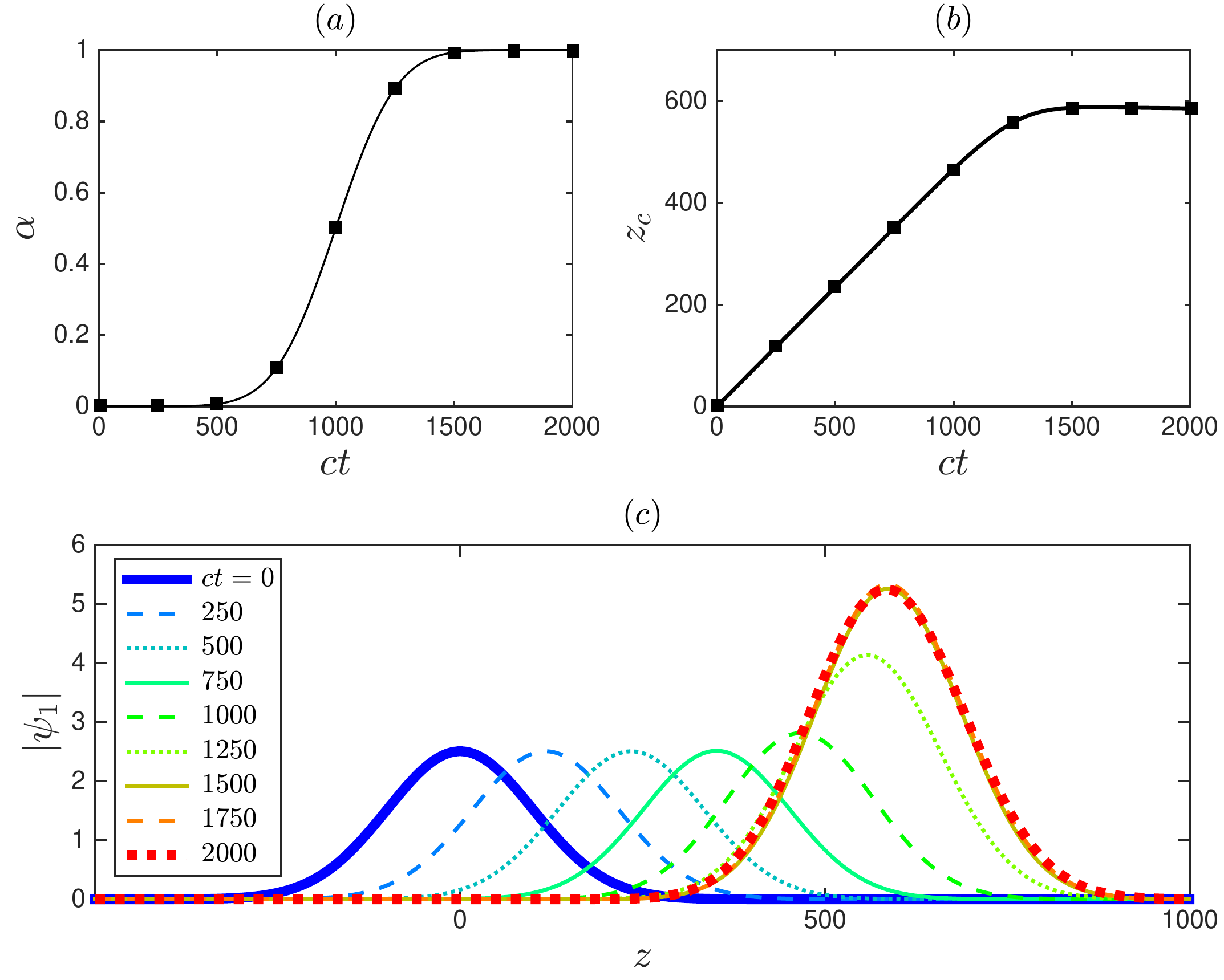}
\caption{(a) Adiabatic change of the gain/loss parameter from $\alpha = 0$ at t = 0 to $\alpha_{EP} = 1$ at final time $t = 2000/c$ ($c$ is the light speed in vacuum) in a system with representative parameters $k = 0.5$, $n_w = 1.6$, and $\delta = 0.5$ (arb. uint). (b) Temporal evolution of the center $z_c$ of the Gaussian pulse, stopping when $\alpha$ reaches the value $\alpha_{EP} = 1$ at EP. The pulse is prepared initially in anti-symmetric mode of the system with no gain/loss with the mean propagation number $\beta = 0.8$ and standard deviation $\sigma = 0.01$. (c) Pulse envelope $|\psi_1|$ in the first WG at times $ct = 0,250,\ldots,2000$, which correspond to squares in the upper figures. At the three latest times, the group speed vanishes and the corresponding graphs collapse to a single curve demonstrating the full-stop of a pulse. A backward change of the gain/loss parameter brings the optical signal to its original mobile form.}
\label{Figure2}
\end{figure}

Conclusions based on the effective model (\ref{eqR3}) are further confirmed with the numerical computation for the full equation (\ref{eqR1}). Here the propagation constants $\beta$ and eigenfunctions $\psi(x)$ are calculated numerically for given frequency $\omega$ by diagonalizing the non-Hermitian Hamiltonian in a matrix representation using a particle in a box basis set. The two modes coalesce at the EP for specific values of $k$ and $\beta$ in the presence of gain and loss, and one can see from Fig.~3(a) that the derivative $d\beta/dk$ becomes infinite at the EP providing the vanishing group velocity $v_g = c\left(d\beta/dk\right)^{-1}$. The corresponding self-orthogonal eigenfunction is given in  panel (b). 

Finally,  Fig.~4 shows the propagation of Gaussian wave packets, comparing the power spectrum $|\phi(x,z,t)|^2$ at the initial time $t = 0$ vs. the final time of $10$ picoseconds. Here the Gaussian solution for a constant gain/loss parameter $\alpha$ is written as $\phi(x,z,t) = A\int
\exp\left(-\frac{(\beta-\beta_0)^2}{2\sigma^2}+i\beta z-i\omega t\right)\psi(x)d\beta$, where both $\omega$ and $\psi(x)$ should be expressed as functions of $\beta$. Note that the Gaussian pulse at the EP contains the contributions from both sides of $\beta_{EP}$, which correspond to two different modes coalescing at the EP in Fig.~3(a). In numerical computations, we used $\beta_0 = 0.851\,\mu m^{-1}$, $\sigma=0.002\,\mu m^{-1}$ and the overall interval $0.845 \le \beta \le 0.857\,\mu m^{-1}$.
In Fig.~4(a), the pulse parameters are chosen exactly at the EP, while figure (b) corresponds to a similar pulse but for the system far from the EP (no gain/loss, $\alpha = 0$). The latter pulse has the large group speed $v_g/c = 0.47$ and the phase speed $v_p/c = 0.82$, demonstrating a considerable dislocation of about $1.4$\,mm in $10$\,ps, while the full-stop is confirmed for the pulse designed at the EP. We used illustrative values of physical parameters in these simulations, which provide a larger frequency window near the EP than those studied experimentally in~\cite{ruter2010observation}. Note that the dispersion curve in Fig.~3(a) exhibits a round shape including also a point with infinite group velocity \cite{wang2000gain,withayachumnankul2010systemized}, where $d\beta/dk = 0$. This point is outside the operation window for our protocol.

\begin{figure}
\centering
\includegraphics[width=0.9\columnwidth]{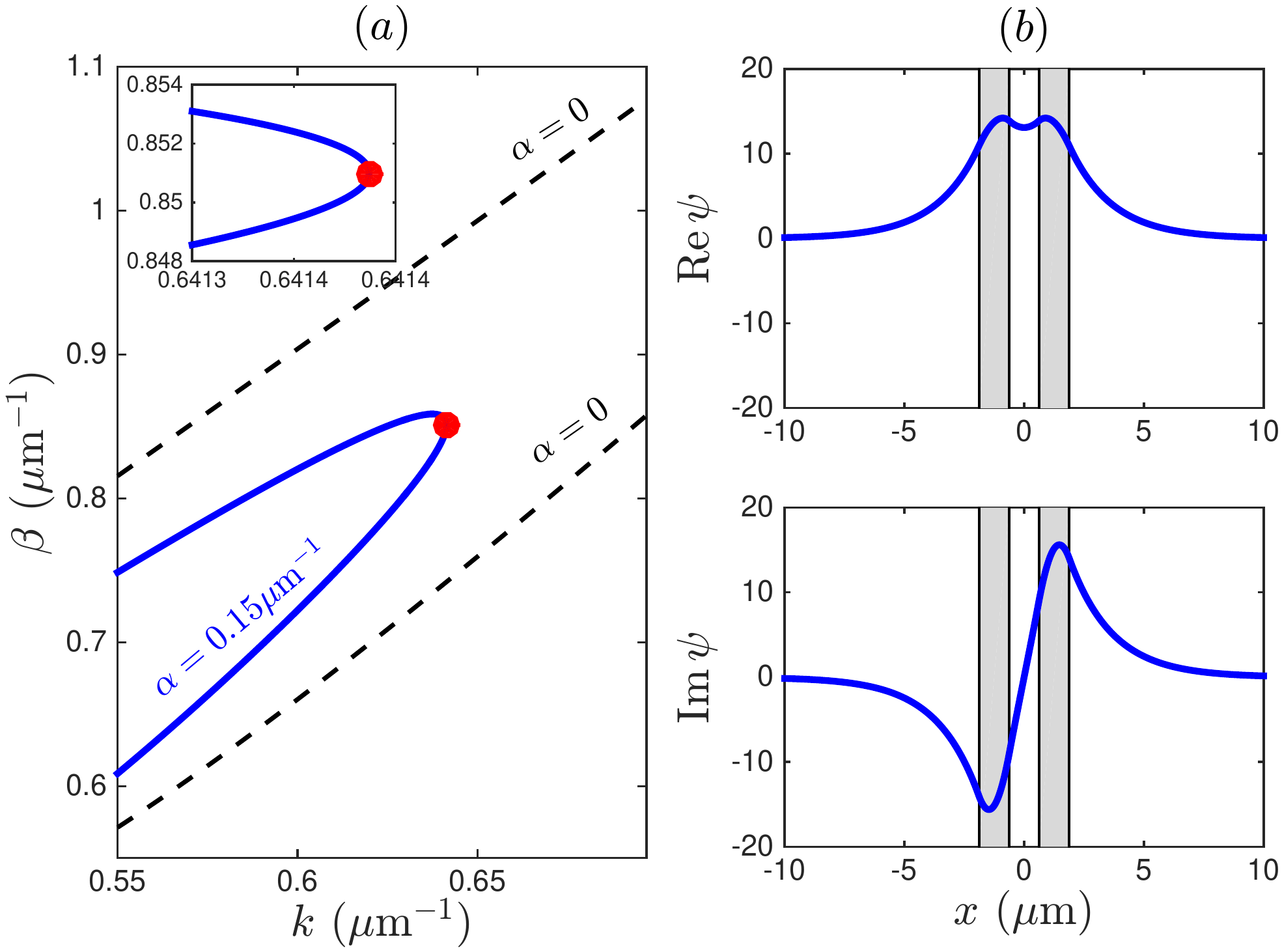}
\caption{(a) The propagation constant $\beta$ as a function of $k=\omega/c$ at two different values of the gain/loss parameter: Hermitian system with $\alpha=0$ (dashed black lines: upper symmetric and lower antisymmetric modes) and non-Hermitian PT-symmetric system with $\alpha=0.15\,\mu m^{-1}$ (solid blue line). Two modes of the PT-symmetric system coalesce at the EP marked with a red circle. Inset shows enlarged vicinity of the EP at $k_{EP}=0.6414\,\mu m^{-1}$ and $\beta_{EP}=0.851\,\mu m^{-1}$. The infinite derivative $d\beta/dk$ at the EP yields the vanishing group velocity $v_g = c\,(d\beta/dk)^{-1}$. 
(b) Real and imaginary parts of the complex eigenfunction for the PT-symmetric system at the EP. This mode satisfies the self-orthogonality condition $\int \psi^2dx = 0$. Grey vertical regions in the background show positions of the two coupled WGs.}
\label{Figure3}
\end{figure}
%(c) The propagation constant $\beta$ as a function of $k$ for the system of Ref.~[9]. This system has the gain/loss parameter $\alpha = 4.04\,cm^{-1}$ and yields the EP at $k_{EP} = 11.6360\,\mu m^{-1}$ and $\beta_{EP} = 26.7697\,\mu m^{-1}$.

\begin{figure}
\centering
(a) PT-symmetric at EP\\[1mm]
\includegraphics[width=0.8\columnwidth]{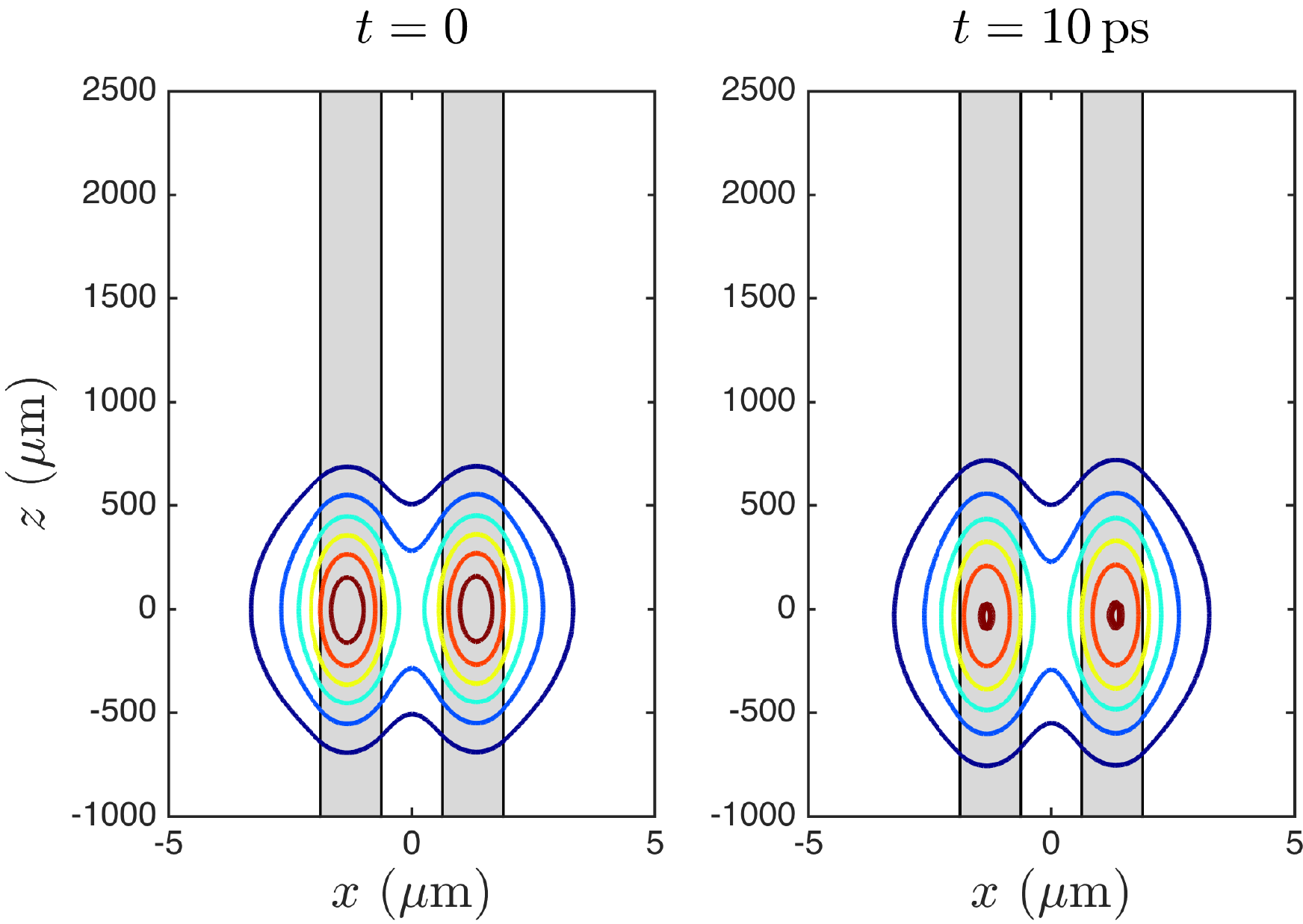}\\[1mm]
(b) Hermitian: no gain/loss\\[1mm]
\includegraphics[width=0.8\columnwidth]{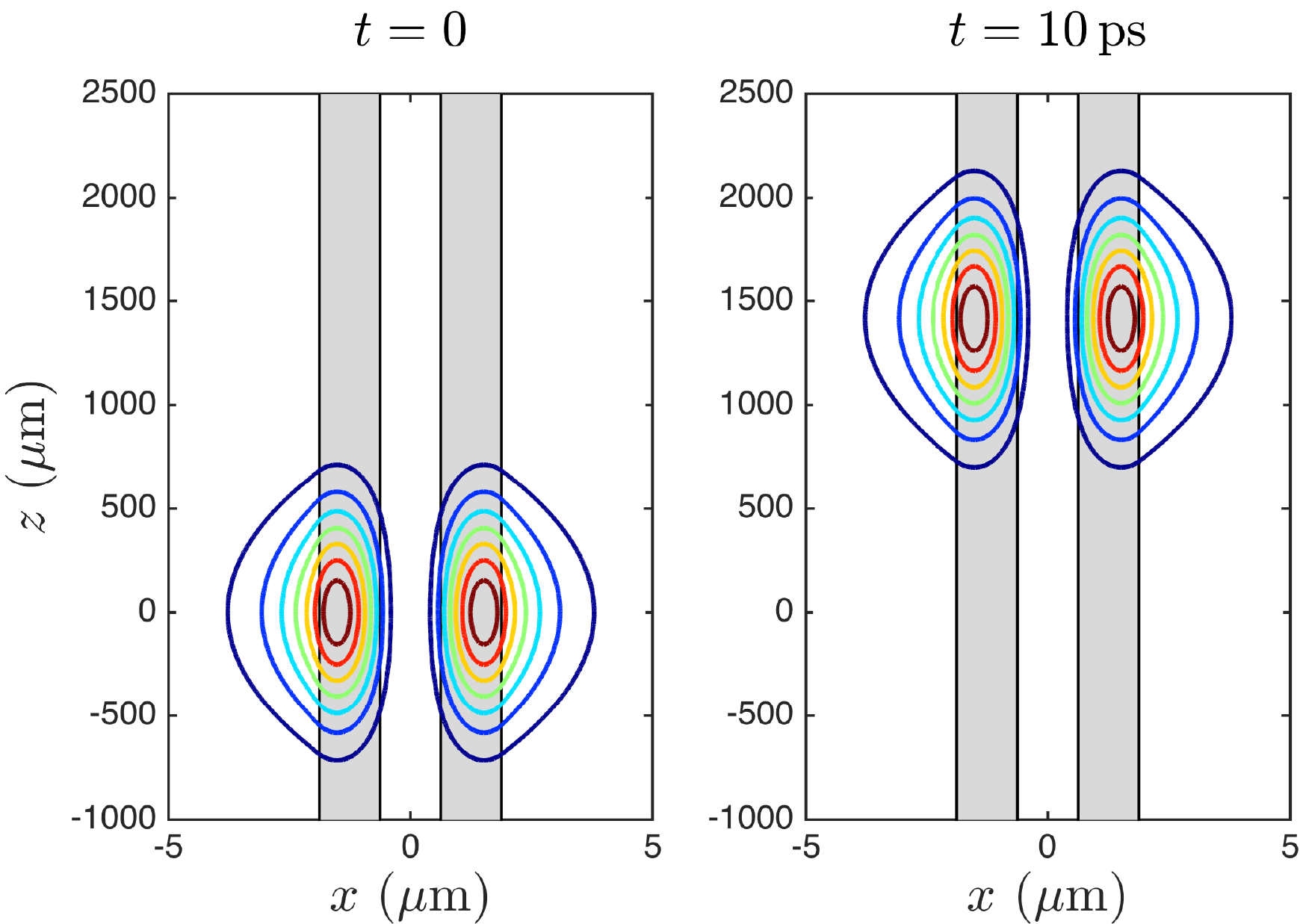}
\caption{Contour plot of the light power $|\phi(x,z,t)|^2$ for the Gaussian wave packets at the initial time $t=0$ and the final time $t=10\,ps$. In both plots the pulse has the mean propagating constant $\beta = 0.851\mu m^{-1}$ with standard deviation $\sigma=0.002\,\mu m^{-1}$. Grey regions in the background show positions of the two coupled WGs. (a) Fully stopped pulse centered exactly at the EP in the PT-symmetric system for $\alpha=0.15\,\mu m^{-1}$. (b) The antisymmetric mode in the Hermitian case with no gain and loss ($\alpha=0$). One can see that the pulse dislocates for about $1.4$\,mm in the Hermitian case, while it does not move at all when prepared at the EP in the PT-symmetric system.  }
\label{Figure4}
\end{figure}

The stopped signal in our system has the phase velocity $v_p/c = 0.75$, which is only weakly affected as the group velocity is reduced to zero under the EP mechanism. Furthermore, the phase speed demonstrates a slight decrease compared to the system with no gain/loss, contrary, e.g., to the well-known relation $v_p \propto 1/v_g$ in special relativity or in optics at the mode-opening.

The major advantages of the proposed protocol is its non-resonant nature, in which the EP can be adjusted to any frequency by tuning the coupling or gain/loss parameters. There is also a benefit of using the time-dependent variation of parameters. In this case an optical pulse is expanded in spatial Fourier modes with the frequency evolving adiabatically along the real dispersion curve in Fig.~3(a). In this way our protocol avoids the instability related to complex modes at frequencies above the EP, as confirmed by our numerical tests in Fig.~2.

We showed that the full-stop of a light pulse is possible at the exceptional point in PT-symmetric coupled waveguides by varying the gain/loss parameter in time. This allows to ``freeze'' and then release the light pulse preserving the carried coherent information. The use of PT-symmetry has practical advantages of keeping a constant intensity of propagating modes and providing a robust protocol that brings the system to the EP. 
The non-resonant mechanism of the proposed phenomenon, due to large flexibility of controlling the EP position, is an important technological advantage, with potential applications for short optical pulses. Specifically, one can engineer this effect in a PT-symmetric system of two waveguide channels. This approach is not limited only to light but can be extended, e.g., to acoustic waves or other fields in physics related to the PT-symmetry.

\vspace{1mm}The authors are grateful to Moti Segev and Meir Orenstein for most helpful discussions. A.A.M. is supported
by the CNPq grant number 302351/2015-9.
N.M. acknowledges the financial support of I-Core: The Israeli Excellence Center "Circle of Light", and of the Israel Science Foundation grant number 1530/15.

\bibliography{refs}

\end{document}